\documentclass[%
reprint,
superscriptaddress,
amsmath,amssymb,
aps,
prab,
floatfix,
]{revtex4-2}

\usepackage{graphicx}
\usepackage{dcolumn}
\usepackage{bm}
\usepackage{xcolor}
\usepackage{tikz}
\usetikzlibrary {datavisualization.formats.functions}

\begin{document}
	
	\newcommand{\I}{\mathrm{i}}
	\newcommand{\E}{\mathrm{e}}
	\newcommand{\D}{\,\mathrm{d}}
	
	\title{Kinematics of ultrarelativistic electrons in periodic fields of moderate strength} 
	\author{Eugene Bulyak}
	\email{eugene.bulyak@desy.de; bulyak@kipt.kharkov.ua}
	
	\affiliation{National Science Center `Kharkiv Institute of Physics and Technology', 1 Academichna str, Kharkiv, Ukraine \\ V.N.~Karazin National University, 4 Svodody sq., Kharkiv, Ukraine}

\date{ \today}
 
 \begin{abstract}
Bright sources of hard x- and gamma-ray electromagnetic radiation are of high demand in physics and technology. Such sources, e.g., x-ray free-electron lasers  (XFELs), undulator- or Compton-based sources of polarized positrons, etc., require intense ultrarelativivstic electron beams with small energy spread. Also each electron  should emit many photons of almost identical spectra per pass through the periodic field of undulator or laser pulse. We develop a kinematic model based on the energy-momentum conservation law and the corpuscular presentation of the periodic-field photons. The model incorporates the mass shift effect in periodic fields. We made an assumption of equality of the average number of photons, which induce the mass shift,  to the Poisson parameter of the coherent field.  The model allows to evaluate the evolution of the spectrum of electron bunch passing through the periodic fields of undulators or laser pulses.  We propose a method of undulator fine tuning aimed at maximization of the spectral brightness.
\end{abstract}
	
	
\maketitle
\section{Introduction}
Bright sources of hard x- and gamma-ray electromagnetic radiation are of high demand in physics and technology. Such sources, e.g., x-ray free-electron lasers  (XFELs), undulator- or Compton-based sources of polarized positrons, etc., require intense ultrarelativivstic electron beams with small energy spread. Also each electron  should emit many photons per pass through the periodic field of undulator or laser pulse: For XFELs, it is typical $\sim 500$\,photons per pass \cite{nakatsutsumi2014}, for  the undulator of International Linear Collider \cite{ilcrdr}, it is projected $50\dots 300$ photons/pass. 

With increase in the electron energy and the frequency of the external driving forces -- undulators of XFELs, lasers of Compton sources -- recoils caused by emission of  quanta induce energy losses.  These losses are random, and affect the energy and the angular spreads of the beam distribution in the phase space. 

Interactions of coherent photons with ultra-relativistic electrons have been extensively studied  both experimentally and theoretically in recent years. 

We constructed a multiphoton model of  interaction of electrons with periodic fields. The model is based on kinematics of the interactions -- the law of momentum-energy preservation.  We consider the periodic field as aggregation of the photon number states, which  are assumed to be distributed according the Poisson law. Relativistic electrons in such field  acquire an additional mass and may interact with an individual number state.  The key assumption of the model is that the mass shift (a classical parameter equal to the average squared transverse momentum acquired in the field) is assumed to be equal to the Poisson parameter $\lambda$ of the laser/undulator photon states distribution.  This interaction results in scattering off the state -- the nonlinear Compton radiation, or generation of higher harmonics in the undulator radiation.   

In the present paper, we are going to study these effects, aimed at evaluation the evolution of the electron spectrum affected by recoils from scattered off such photon states. 
We show, that the interactions of  electrons with the coherent field are governed by three parameters: (i) the electron energy, (ii) the photon energy, and (iii) the average number of the photons per the laser/undulator wavelength. 

The paper is organized as following: The second section contains a formulation of the suggested kinematic model. In the third section, there are presented considerations of equivalence the classical field of undulators and the quantum coherent  field of laser pulse.  The fourth section describes electron bunch kinetics in strong periodic fields. The paper is ended with summary of results and discussion. 

\section{Problem setup}
The chosen model  aims to study the electron beam kinetics in the sources of hard x-ray and gamma-ray radiation. 
Specific to these sources are the ultrarelativistic electrons that are emitting photons with energy much smaller than that of the electrons. This condition enables to acquire high intensity photon beams since each electron may radiate a number of photons with near identical spectrum.    

Therefore, we avoid some extremal parameters of electron-to-field interaction, such as extreme field strength and extreme energy of the income photons , see a classification of the Compton-like scatterings in \cite{PhysRevAccelBeams.27.080701}. 

We use the natural system of  units for particle and atomic physics, $\hbar = c = m_e = 1$, where appropriate. Also, the energy of photons will be reduced to the electron rest energy:
$\Omega = \hbar \Omega_\text{las}/(m_e c^2)$ and the like for scattered off photon $\omega$.
 
In this system, the reduced four-momentum vectors [contravariant, the metric signature $\left( +,-,-,-\right)$] of the electron and the photon are :
\begin{align*}
	P_\gamma &=[\gamma ,\sqrt{\gamma^2-\zeta}, 0,0]\; ; \\
	P_{\gamma'} &=[\gamma' ,\sqrt{\gamma^2-\zeta}, 0,0]\; ; \\
	P_\Omega &= [\Omega,-\Omega,0,0]\; ;\\
	P_\omega &= [\omega,\omega,0,0]\; ,
\end{align*}
where $\gamma,\gamma'$ are the initial and final Lorentz-factors of the electron, $\Omega, \omega $ are the initial and the scattered reduced energies of the photon [equivalent Lorentz factors, $\hbar \omega / (m_e c^2)$]. The dimensionless parameter $\zeta\ge 1$ corresponds to the electron mass-shift  in the intense periodic field, see \cite{PhysRevLett.109.100402}:
\[
m^2_e c^4\Rightarrow \zeta m^2_e c^4\; .
\]   

Kinematic relations (the energy and momentum conservation laws)  $P_\gamma + P_\Omega =  P_{\gamma'} + P_\omega$ enable one to derive the maximum energy of scattered (undergone recoil)  photons: 
\begin{equation}\label{eq:maxomega}
\omega_m = \frac{\Omega\left[\left( \gamma+\Omega+\sqrt{\gamma^2-\zeta}\right)^2-\Omega^2 \right]}{4\Omega(\gamma+\Omega)+\zeta}
\end{equation}

It is worth to note, that the relation \eqref{eq:maxomega}  is exact, only restricted to  $\gamma^2\ge \zeta$: the classical Compton formula,  $\gamma= \zeta=1$ is a limiting case.

For the considered case of the ultrarelativistic electrons $\gamma\gg\Omega$,  Eq.~\eqref{eq:maxomega} is reduced to the known ones (see, e.g., \cite{abramowicz21,PhysRevAccelBeams.27.080701}):
\begin{equation}\label{eq:maxom}
\omega_m\approx \frac{4\Omega\gamma^2}{4\gamma\Omega+\zeta} \; .
\end{equation}
The first term in the denominator, $4 \gamma\Omega$,  is equal to the income photon energy in the electron rest frame.

The final energy of the electron that emitted a photon on the Compton edge is
\[
\gamma' = \gamma+\Omega-\omega \; . 
\]

\section{Equivalence of classical field energy and photons density}
Parameter $\zeta $, used in the previous section,  is connected to the field strength as $\zeta := 1+\xi^2$, where
the laser field strength parameter $\xi $ -- the classical nonlinearity parameter, see \cite{abramowicz2023tech} -- is : 
\[
\xi: = \left<|E|\right>  \frac{|e|}{m_e c\,\Omega_\text{las} }
\]
where $\left<|E|\right>$ is the averaged over  period module of the electric field strength, $e, m_e$ are the charge and the mass of electron, resp., $c$ is the velocity of  light, and $\Omega_\text{las}$ is the angular frequency of the laser field.

Formal presentation of the parameter $\lambda:=\xi^2 $ as the number of equivalent photons per their wavelength through the electric field squared needs to be clarified:  The \emph{average} over period electric field strength, $E^2$ in
\[
 \left< \xi^2 \right> = \left<E^2\right>  \left(\frac{e}{mc\Omega } \right)^2\; ,
\]
is equal to its maximum value for the circularly polarized wave. For the linearly polarized wave, the average photons density is $\lambda = \xi^2/2$.  

Employing the physical constants, the fine-structure constant $\alpha$ and the Compton electron wave length $\lambdabar_C$,  
\[
\alpha = \frac{e^2}{\hbar c}\; ;\quad \lambdabar_C = \frac{\hbar}{mc}\; ,
\]
and taking into account the energy density (CGS system of units)
\[
w = \frac{1}{8\pi}\left(E^2+H^2\right)
\]
($|E|=|H|$ in flat wave) we finally obtain
\[
\xi^2=2\alpha \lambdabar_C^2 \lambda_\text{las}\rho_q \; .
\]

Here,  $\rho_q =E^2/4\pi \hbar\Omega_\text{las}$ is the density of effective laser photons. Therefore, $\xi^2 $ is equal to the mean number of laser photons contained in the volume
\[
V= 2\alpha\lambdabar_C^2 \lambda_\text{las}\; .
\]

Numerical value for  the `effective volume':
\[
V = 2.18\times 10^{-27}\lambda [\text{m}]\, \text{m}^{3} = 2.18\,\lambda[\text{nm}]\, \text{nm}^{3}\; .
\]

Recollecting  the Thomson cross section, $\sigma_T = 8\pi r_e^2/3 = 8\pi \alpha ^2 \lambdabar_C ^2/3$, the condition $\lambda = \xi^2 = 1$ may be presented as a scattering of one photon over
\[
\frac{3}{4\pi\alpha}\approx 33\,\text{periods} \; .
\] 
  
\subsection{Quantization of the classical periodic electromagnetic field}
Interaction of the ultrarelativistic electron $\gamma \gg 1$ with the periodic magnetic field of  undulator magnet is similar to the interaction with the  flat  EM wave. In the electron rest frame, the difference between the actual electric field $E_w$ and that of the flat wave $E$ is significantly reduced:  
\[
\left(E^2_w - E^2 \right)/E^2_w \sim \gamma^{-2}\; ,
\]
as a consequence of a Lorentz-invariant scalar $\left(E^2-H^2\right)=\text{const}$. 

The role of the laser $\xi^2$ parameter is played by the undulator $K^2$ parameter [$K:= e B_0\lambda_\mathrm{u} / (2\pi m c )$, see \cite{kincaid77}]. There is no electrical field in the lab frame, therefore the energy density is half of the equivalent  wave;  on the other hand, the effective energy of the quantum is half of the wave's. [For ultrarelativistic electrons,  the undulator field may be treated as the flat wave with the half of frequency: $\Omega_u = c/(2\lambdabar_u)$.]

Thus, $\lambda:=\xi^2$ is equal to the number of photons that `dressed up' the electron passing though an oscillating field.

\subsection{Fundamental assumption: coherent field} 
In the photon number presentation, a coherent state is defined by the Poisson-weighted sum of the number units, see, e.g.,  \cite{glauber63,fox06}, each containing  $k$ photons: 
\[
p_k = \E^{-\lambda} \lambda^k / k!\; ,
\]
where $p_k $ is the probability of having a unit state containing exactly $k$ photons, and $\lambda $ is the mean number of photons. 

Let us consider a mass-shifted (dressed) electron scattered off a certain number unit `converting' it into a single photon (the classical analog of harmonics): $\gamma + k\Omega\Rightarrow \gamma' +\omega$. Its maximum energy, c.f., Eq.~\eqref{eq:maxom}, is
\begin{equation}\label{eq:maxomk}
	\omega_{k}^{max}\approx \frac{4 k \Omega\gamma^2}{4\gamma k\Omega+\zeta} \; .
\end{equation}

The key assumption is: the mean number of photons, that the electron coupled with, equals to the magnitude of the mass shift :
\begin{equation}\label{eq:kmeqlambda}
\overline{k} = \lambda\; .
\end{equation}

With substitution $\Omega\Rightarrow k\Omega $ into Eq.~\eqref{eq:maxom},  we deduce  the maximum photon energy in $k$th harmonic:
\begin{equation}\label{eq:maxomk}
	\omega_{k}^{max}\approx \frac{4 k \Omega\gamma^2}{4k\gamma\Omega+1+\lambda } \; .
\end{equation}

For a given spectrum of each harmonic $f_k(\omega)$, the aggregate spectrum reads:
\[
F_\lambda(\omega ) = \sum_{k=1}^{\infty}\frac{\E^{-\lambda}\lambda^{k-1}}{(k-1)!} f_k(\omega;\lambda)\; .
\]

The quasi-Thomson limit of small recoil, $4 \gamma \Omega\ll 1+\lambda $, which is specific to the undulator-  and Compton sources of  x-ray  radiation, each harmonic is symmetric around its mean frequency $\overline{\omega_k}$, see \cite{kincaid77} and Appendix~\ref{apa}. In this practical approximation, the mean energy in the aggregate spectrum is \emph{independent} from the field strength:
\begin{equation}\label{eq:meanom}
\overline{\omega}:=\int \omega F_\lambda(\omega ) \D\omega = \omega^
{max}_1(\lambda\to 0)/2.
\end{equation}

It should be noted, that for a  very high harmonic number $k\to \infty$, the approximation of negligible recoil does not hold. But for the Poisson distribution, the contribution of these harmonics to the aggregate spectrum is exponentially small at $k\gg \lambda $. 

\section{Electron bunch kinetics}
Under the assumption of independently emitted photons with the  identical spectrum, and thus independent recoils undergone by the electron, evolution of the energy spectrum of the bunch -- its characteristic function -- is governed by the formula, \cite{bulyak17b,bulyak18,bulyak19a}:
\begin{equation}\label{eq:BSH}
	\hat{f} = \hat{f}_0\exp[ x(\check{w} - 1)] =:\hat{f}_0 \hat{S}
\end{equation} 
where $\hat{f}$ is the Fourier transform of the probability density function (PDF) -- the characteristic function for the energy spectrum -- for the electron bunch, $\check{w}$ is the inverse Fourier transform of the recoil spectrum, $x$ is the ensemble average number of recoils from the front end of the field, $f_0=f(x=0)$. 

Therefore, for the evaluation of the  evolution of bunch spectrum, it is sufficient to evaluate the (inverse) Fourier transform of the recoil spectrum.

The spectrum $f(\gamma;x,w)$ is the convolution of the initial spectrum $f_0(\gamma )$ with the struggling function $S(x;w)$.  The straggling function -- distribution density of fluctuations -- is determined solely by the ensemble-average number of recoils having undergone by the particle since entering the field,  and the spectrum of the recoil. 

\subsection{ Evolution of spectrum} 
Equation \eqref{eq:BSH} provides comprehensive description of the electron bunch kinetics that undergoes random independent identically distributed energy losses.
For the Thompson scattering, where the spectrum has a compact support, $\omega\in (0,\omega^{max}]$, we can make a few estimations of the spectrum evolution  dependence on the average number of scattering:
\begin{itemize}
	\item [$x\lesssim 1$] Contribution of the initial distribution decays as $f_0 \exp(-x)$ with addition of a tail directed toward low energy.  The tail is formed from the electrons subjected to one ($\propto x$), two ($\propto x^2/2$), etc. convolutions of the recoil spectrum with the initial distribution. 
	The spectrum may have a number of local  maximums.
	\item[$x\lesssim 10 $] The spectrum evolves toward a unimodal one (single maximum), which may be approximated with a stable univariate distribution \cite{nolan:book1}, the stability parameter $\alpha (x) $ increases from one (ballistic diffusion), and asymptotically approaches two (the normal diffusion).
	\item[$x\to \infty$] Quasi-normal diffusion, the distribution is Gaussian with truncated high-energy tail, $f(\gamma > \gamma_0^{max})=0$.   
\end{itemize}

Equation~\eqref{eq:BSH} allows us to derive the moments of the spectrum:
\begin{align}\label{eq:moments}
	\overline{\gamma}(x) &= \overline{\gamma_0} - x\,\overline{\omega}\; ; \nonumber \\
	\mathrm{Var}[\gamma ] (x) &=\overline{(\gamma-\overline{\gamma})^2}=\mathrm{Var}[\gamma_0 ] + x\, \overline{\omega^2}\; ; \nonumber
	\intertext{in general}
	M_n^{\text{cnt}}[f] &=M_n^\text{cnt}[f_0]+(-1)^n x  M_n^{\text{raw}}[w]\,; \quad n>1 ,  
\end{align}
where $ M_n^{\text{cnt}}, M_n^{\text{raw}}$ are the $n$th centered and raw  moments, respectively.

All the moments of the electron spectrum are linear on the number of scatterings $x$, which in turn is linear on  the number of the laser photons per the wavelength $\lambda$. Since the average energy loss per interaction is independent of the number of photons $\lambda $, see \eqref{eq:meanom}, the evolution of the straggling function for a given recoil spectrum is determined only by the number of recoils $x$.

The moments of PDF  provide adequate description of the evolution for a big number of recoils, $x\gg 1$: The diffusion approximates the normal one, the scale parameter (width of spectrum ) $c\sim x^{1/2}$, and the odd moments become negligible.

At  $x\gtrsim 1$, the evolution of  the spectrum differs from the normal. The width of the spectrum  depends on the number of recoils  (see \cite{bulyak19a}) as:
\begin{equation} \label{eq:cspec}
c (x)= \left[ x m_\alpha[w]\right]^{1/\alpha }\; , 
\end{equation}
where
\[
\alpha(x) = \left. \frac{s D_s \Re [\check{w}]}{1-\Re [\check{w}]}\right|_{s=s_*(x)} \;,
\]
$s_*(x)$ is a root of equation
\[
\Re [x\left( 1-\check{w}(s_*)\right)] = 1\; ,
\]
and $m_\alpha[w]$ is the fractional moment of the recoil spectrum:
\begin{equation}\label{eq:frmom}
m_\alpha[w] = \int_R \omega^\alpha w(\omega)\D \omega\; ; \qquad \omega > 0\; ,
\end{equation}
$D_s\Re[\check{w}] $ indicates the partial derivative with respect to $s$ of  the real part of the inverse Fourier transform of recoil spectrum.

\subsection{Aggregate spectrum of recoils}
The `realistic' aggregate spectrum of nonlinear Compton radiation (if the components do not differ significantly from the Thomson's) may be approximated as composition of rectangles with the Poisson weight:
\begin{equation} \label{eq:spectra}
	w(\omega; \gamma , \Omega , \lambda) = \sum\limits_{k=1}^\infty {P_{k-1}[\lambda ]\frac{\Theta[\omega,\omega_k^{max}-\omega]}{\omega_k^{max}}}\; ,
\end{equation}
where $\Theta(\cdot, \cdot )$ is the Heaviside 2D step function. 

A closed form of the spectrum \eqref{eq:spectra} read
\begin{align}\label{eq:aggrsp}
w(\omega; \gamma , \Omega , \lambda) &=4\gamma^2\Omega\frac{(\lambda+1)} {\lambda} \left[1-\mathrm{Q}(\lceil(\lambda+1)\omega\rceil,\lambda)\right]\nonumber \\ 
& := 4\gamma^2\Omega \times G(\omega,\lambda)\; 
\end{align} 
where $\lceil x\rceil]$ gives the ceiling (the smallest integer greater than or equal to $x$), $ \mathrm{Q}$ is the incomplete gamma function.

The aggregate recoil spectrum  may be factorized into  the relativistic Thomson factor $4\gamma^2\Omega$  and a spectral function $G(\omega,\lambda)$ independent on $\gamma$ and $\Omega $, [$\mathrm{Q}(\cdot,\cdot)$ is the regularized incomplete gamma function].  

\begin{figure} 
	\includegraphics[width=\columnwidth]{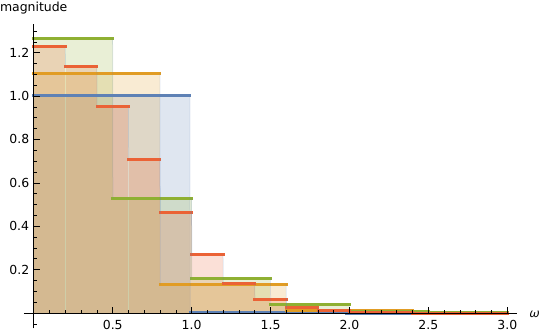} \\
	\includegraphics[width=\columnwidth]{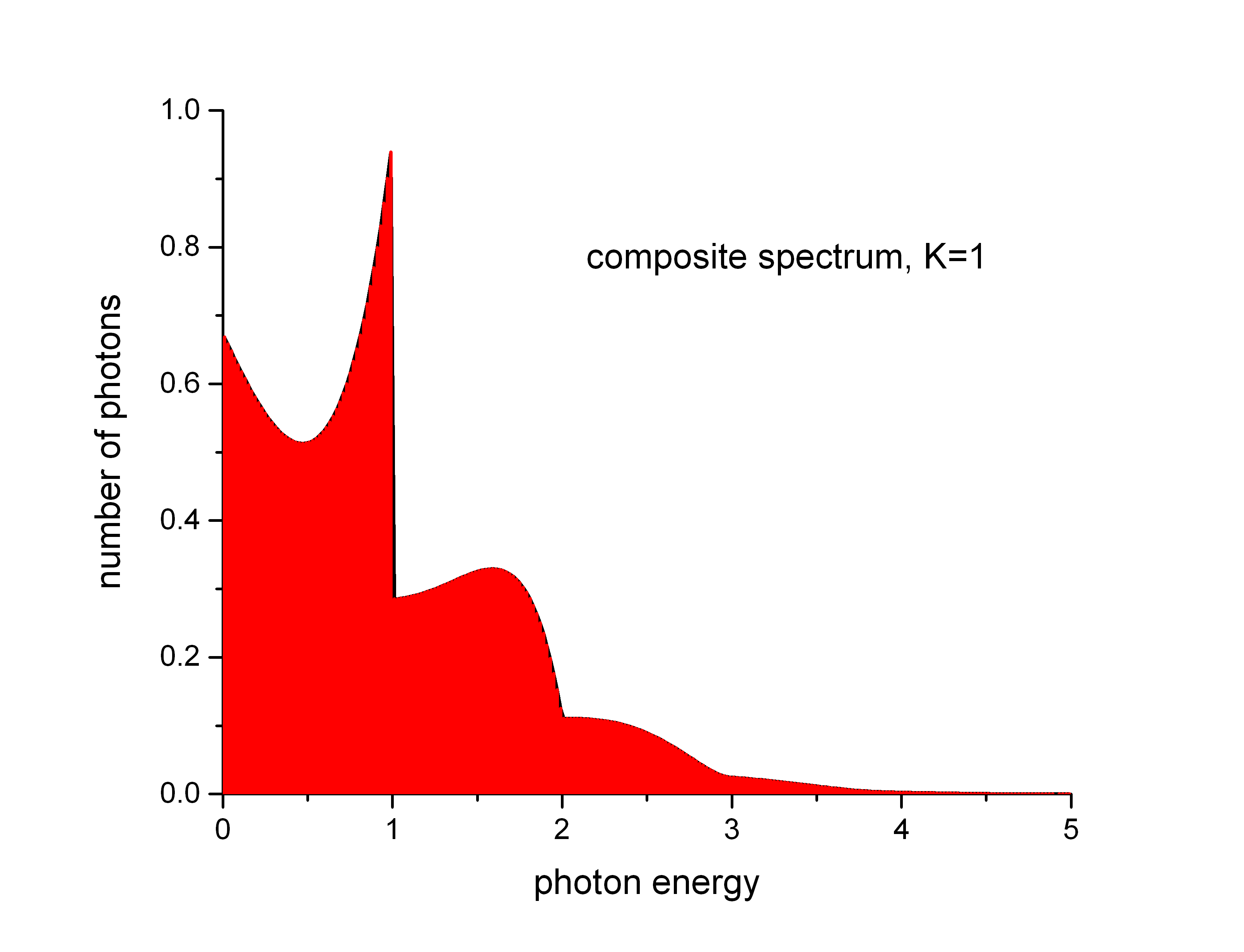}
	\caption{Top panel: Spectral function $ Q(\omega,\lambda)$  for $\lambda = 0.01,0.25,1,4$ (blue, orange, green, and red, resp.). 
	Bottom panel: Superposition of the first five harmonics \cite{kincaid77} with the Poisson weight, $K^2 = \lambda = 1$.  Photon's energy normalized to the ensemble average.	\label{fig:foursp1}}
\end{figure}

Figure \ref{fig:foursp1}  presents the function  $G(\omega,\lambda)$
for $\omega_1^{max}(\lambda\to 0) =1 $, and for a range of the field strengths $\lambda $.
 
As seen from Fig.~\ref{fig:foursp1}, all the harmonics shrink with increase of $\lambda $, while the amplitudes of higher harmonics increases.  These simplified spectra look like a realistic one, see Fig.~\ref{fig:foursp1} bottom panel. 

\subsection{From spectrum of recoils to spectrum of electrons}
The inverse Fourier transform ($\omega\to s$) of the aggregate recoil spectrum $G(\omega,\lambda)$ has a form:
\begin{subequations}\label{eq:ifw}
	\begin{align}
		\check{w}(s;\lambda)&= \frac{\I }{\lambda \phi } \left\{ 1-\exp\left[ \lambda \left(\exp\left( \I\phi \right) -1\right)\right] \right\} \label{eq:ifwt}\\
		& = \frac{ \I}{\lambda\phi} \left\{ 1-\exp\left[-\lambda(1-\cos(\phi))\right] \cos(\lambda\sin(\phi))\right\} \label{eq:ifwim}\\ 
		&+ \frac{1}{\lambda\phi}\exp\left[-\lambda(1-\cos(\phi))\right] \sin\left[ \lambda\sin(\phi)\right] \label{eq:ifwre}\; ,
	\end{align}
\end{subequations}
where $\phi := 2\pi s/(1+\lambda)$ .

\begin{figure} 
	\includegraphics[width=\columnwidth]{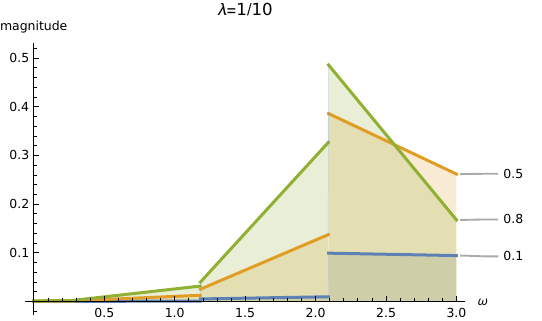} \\
	\vspace{20pt}
	\includegraphics[width=\columnwidth]{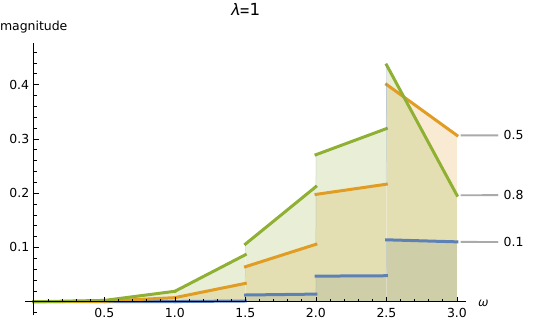} \\
	\vspace{20pt}
	\includegraphics[width=\columnwidth]{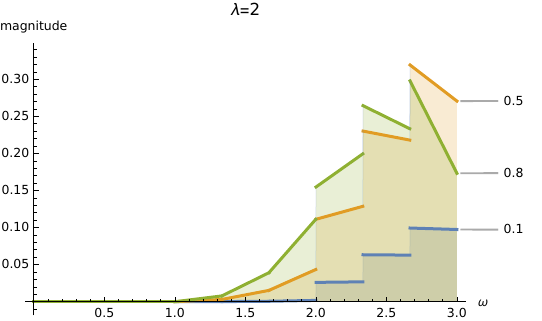}
	\caption{Electrons' spectra for different $\lambda $ at $x=0.1, 0.5, 0.8$, resp., initial delta function not shown. \label{fig:lambda}}
\end{figure}

The electron bunch spectrum -- i.e., the probability density function (PDF) -- under the assumption  of  initial delta function $f_0 = \delta(\gamma-\gamma_0)$ is presented in Fig.~\ref{fig:lambda}.

The initial electron spectrum decays with $x$ as $\sim\exp (-x ) $. The mode (maximum) position  moves to the left jump-like, shifting by the width of the first harmonic.  The case $\lambda = 1$ is characterized by the maximum tail `weight'. 

The second stage of the spectrum evolution, $x\lesssim 10 $, may be approximated with $\alpha$-stable distribution, where the stability parameter is a function of $x$. 

The scale $c$ -- the width of spectrum -- evolves as
\begin{equation}\label{eq:scale} 
c (x)= \left[ x m_\alpha[w]\right]^{1/\alpha } \; ,
\end{equation}
where $m_\alpha $ is a fractional moment of the recoil PDF \eqref{eq:frmom}.
%

\begin{figure} 
	\includegraphics[width=\columnwidth]{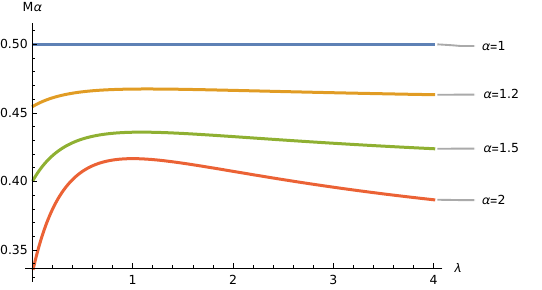}
	\caption{Moments $M_\alpha (\lambda )$ for different $\lambda$.  \label{fig:malpha}}
\end{figure}

The fractional moments of the considered spectrum are functions of $\lambda $, see Fig.~\ref{fig:malpha}.

With an increase of the stability parameter $\alpha$, the scale $c$ -- width of PDF -- builds up slower with number of the emitted photons. 

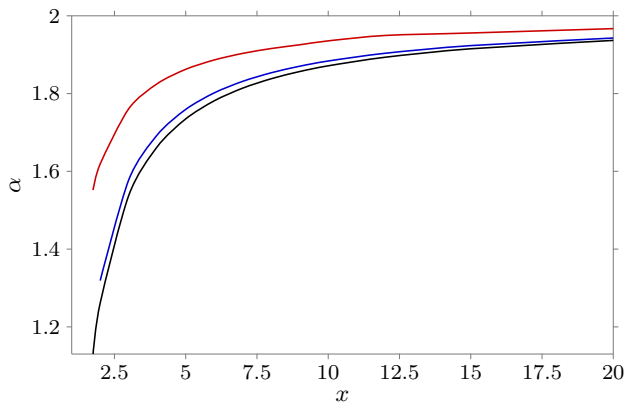
\begin{figure}
	\begin{tikzpicture} [scale=0.9] 
	\datavisualization [scientific axes, x axis = {
		include value=1,
		length=8 cm,label={$x$ }}, y axis = {length=5 cm,include value=2.0,label={$\alpha$}}, visualize as smooth line/.list={tot,pos,neg},
	style sheet=strong colors]	
	
	data [set=tot] {
		x, y
		1.75, 1.13017
		2, 1.26139
		3, 1.53774
		4, 1.66309 
		5, 1.73485
		6, 1.78137
		7,  1.81399
		8, 1.83814
		9, 1.85669
		10, 1.87139
		12, 1.89344
		15, 1.91518
		20, 1.93669
	}
	data [set=pos] {
		x, y
		1.75, 1.55177 
		2, 1.6191
		3, 1.76087
		4, 1.82536
		5, 1.86237
		6, 1.88641
		7, 1.9033
		8, 1.9158
		9, 1.92545
		10, 1.93555
		12, 1.94945
		15, 1.9558
		20, 1.96699
	}
	data [set=neg] {
		x, y
		2,1.31903  
		3,1.57793  
		4, 1.69353  
		5, 1.75928   
		6, 1.80176  
		7, 1.83148  
		8, 1.85345  
		9, 1.87034  
		10, 1.88374  
		12,1.90365   
		15, 1.92334  
		20, 1.94282  
	};
\end{tikzpicture}

	\caption{Stability parameter $\alpha (x; \lambda )$ vs. $x$ for $\lambda = 0.01$ (red),  $\lambda = 1$ (black), and $\lambda = 2$ (blue).  \label{fig:alphax}}
\end{figure}

\begin{figure}
	\begin{tikzpicture} [scale=0.9] 
	\datavisualization [scientific axes, x axis = {
		include value=1,
		length=8 cm,label={$x$ }}, 
	y axis = {include value= 0.5,
		length=5 cm,label={scale}}, 
	visualize as smooth line/.list={tot,pos},
style sheet=strong colors]	

data [set=tot] {
	x, y
	1.75, 0.853207 
	2,  0.935148  
	3, 1.18576   
	4, 1.37715   
	5, 1.53884    
	6, 1.68181   
	7,  1.81154   
	8,  1.93125 
	9,  2.04306 
	10, 2.1484   
	12, 2.3433   
	15, 2.60635   
	20, 2.99061   
}
data [set=pos] {
	x, y
	1.75, 0.853911 
	2, 0.91287 
	3, 1.11803 
	4, 1.29099 
	5, 1.44337 
	6, 1.58114 
	7, 1.70782 
	8, 1.82574 
	9, 1.93649 
	10, 2.04124 
	12, 2.23606 
	15, 2.5 
	20, 2.88675 
};
\end{tikzpicture}

	\caption{The scale  vs number of photons scattered off for $ \lambda = 0.01$ (red), and $\lambda = 1$ (black).  \label{fig:salpha}}
\end{figure}
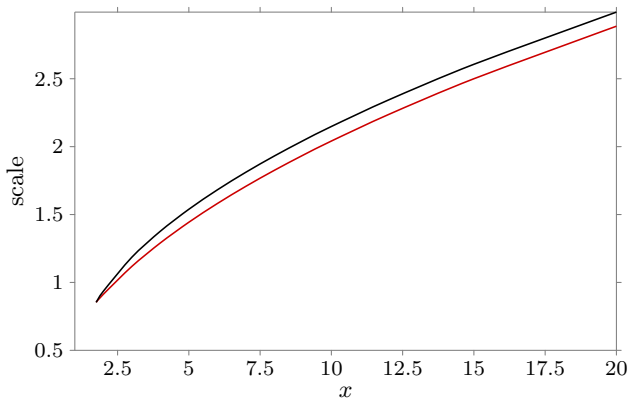

Dependencies of the stability parameter on the number of recoils for different $\lambda $ are presented in Fig.~\ref{fig:alphax}. Corresponding growth of the scale parameter with $x$ is presented in Fig.~\ref{fig:salpha}.

In the intermediate range of recoils, $1\lesssim x \lesssim 10$, the width of the spectrum is increasing faster than predicted for the normal diffusion. The rate of the increase 
depends nonlinearly on the photons density $\lambda $ with its  maximum around $\lambda = 1$.
 
For the big number of scatterings, $x\gtrsim 20 $, when the electron spectrum is not too different from the normal distribution, $\alpha \lesssim 2$,  essential information can be deduced from the moments. The normalized central  $n$th moment ($n>2$)  decreases with the increase of $x$:
\[
\mu_n\equiv \frac{M_n}{M_2^{n/2}}\propto \frac{x}{x^{n/2}}=x^{1-n/2}\; ,
\]
where $M_n$ is the central $n$-moment, c.f. Eq.~\eqref{eq:moments}. 

When the stability parameter approaches the `normal diffusion' magnitude of $\alpha \lesssim 2$ (still remaining below it), the skewness $\mathrm{Sk} [w] > 0$ -- the third central moment -- remains positive and increases linearly with $x$. We can derive practical information about the mode of the distribution, i.e.,  the coordinate of the maximum. Making use of Pearson’s skewness for a distribution close to the normal, see, e.g. \cite{kenney62,bulyak19a},
the maximum of spectrum (mode) is shifted from the mean position to
\[
\gamma_\text{mode} = \overline{\gamma} + \frac{1}{3} \mu_3 M_2^{1/2} \approx \overline{\gamma} +\kappa \frac{1}{2} \frac{M_3^{\text{(raw)}}[w]}{M_2^{\text{(raw)}}[w]}\; ,
\]
where $\kappa:=4\gamma^2\Omega$.
For $x\gg 1$, the shift is \emph{independent} of the number of emitted photons . 

The `rectangular' model yields the mode position:
\begin{align*}
\gamma_\text{mode} &= \overline{\gamma} + \kappa L(\lambda)\; ; \\
L(\lambda) &:= \frac{3(1+7\lambda+6\lambda^2 +\lambda^3)} {8(1+\lambda)(1+3\lambda+\lambda^2)} \; .
\end{align*} 

\begin{figure} 
	\includegraphics[width=\columnwidth]{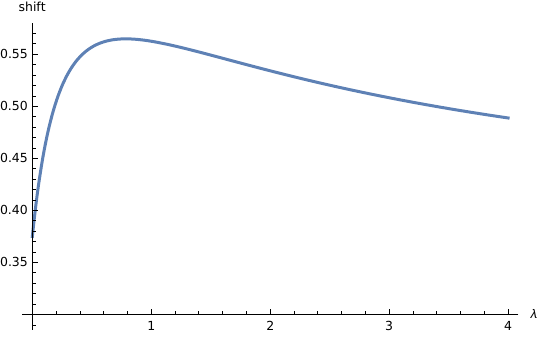}
	\caption{Shift of the mode from the mean vs. $\lambda$.  \label{fig:shift}}
\end{figure}

The function $L(\lambda)$ is plotted in Fig.~\ref{fig:shift}, the maximum shift is attained  at $\lambda_*\approx 0.786 $.

\section{ Results and Discussion}
The energy spectrum of the radiating electrons weakly depends on the driving field magnitude (the higher harmonics composition), which enables us to optimize tapering of the FEL undulators with adjustment to the mode of electron spectrum.

\subsection{Results}

(1)  The proposed kinematic model  comprises of the mass shift of the ultrarelativistic electrons passing through periodical (electro-)magnetic coherent fields. 

(2) The key assumption consists of equality of the mass shift parameter to the average number of photons  that 
the electron interacts.

(3) The considered coherent electromagnetic field has the Poisson distribution of the  photon number states, the Poisson parameter -- ensemble average number of photons the electron interacted with -- equal to that of the mass shift factor.

(4) The model yields well-known dependence of the radiation power on the field strength squared. The power is independent of the harmonics composition.

(5)  A simplified `rectangular' model was developed aimed at deriving analytics of evolution of the bunch kinetics as a function of the field strength. The model is  based on a given cross section of the Compton scattering, the Poisson distribution of the photon number states, and the Poisson-subordinate process of energy losses of electrons  caused by the nonlinear quasi-Thompson scattering.  
 
(6) The most `harmful' field strength -- the widest recoil spectrum and thus the fastest increase of the bunch spectrum with number of emitted photons -- corresponds to one `laser' photon over the period: $\lambda = 1$. 

\subsection{Discussion}

The developed  kinematic model is in  agreement with a few  known results:
First, the $k$-harmonic Compton edge \eqref{eq:maxomk} is identical to that derived for the `nonlinear  QED' case, see \cite{abramowicz21,Harvey_2009}.
Second, the Poisson distribution of photons over the harmonics as well as equality of the mean energy of photon in nonlinear spectrum to that in the linear spectrum ($\lambda\to 0$) agrees well with the classical model of the undulator radiation \cite{kincaid77}.
Third, the Poisson distribution of number of photons in each harmonic together with $k (1+\lambda)^{-1}$ dependence of the mean photon energy in $k$th harmonic yields the classical dependence of the radiation power proportional to the magnetic-field strength squared. 
The latter may validate the effect of the electron mass shift in periodic fields, which has  not yet been proven experimentally,  see \cite{PhysRevLett.109.100402}.

The developed approach will be a useful basis for simulating codes, that  model evolution of the spectrum of electrons passing through nonlinear undulators or laser pulses.   (A linear code was already  realized,  see \cite{bulyak17b}.)  

The above presented  analysis of multiphoton-interacted electron beam  kinetics allows us to fine tune the undulator. With adjustment of the resonant frequency of undulator to the maximum (mode) of spectrum allows to maximize the efficiency of its operation, see  \cite{wang09j,mak16}. An example of tuning a (hypothetic) undulator is presented in Appendix~\ref{apb}.  

The width of electron bunch spectrum increases linearly over the initial section of Free Electron Lasers where the coherent microbunches formed. With increase in energy (aimed at attain higher energy photons) the length of formation increased $\propto \gamma $ while the width of the spectrum is proportional to $\gamma ^2$ that limits the formation of coherent microbunches.  

\begin{acknowledgments}
The author would like to thank Drs. Svitozar Serkez, Gianluca Aldo Geloni, and Serguei Molodtsov of EuXFEL for fruitful cooperation and  interest. Multi-photon processes were extensively discussed with the  late Prof. Nikolay Shul'ga of NSC KIPT.  Computations were done with the \textsc{Wolfram Mathematica Cloud} package \cite{Mathematica}.
\end{acknowledgments}

\providecommand{\noopsort}[1]{}\providecommand{\singleletter}[1]{#1}%

\appendix

\section{Kincaid's classical spectra, according to \protect\cite{kincaid77}\label{apa}}

Let us start from formula (25) in \cite{kincaid77} for the frequency spectrum $I(\omega)$:
\begin{align}\label{eq:Kin25}
	I(\omega)&=\frac{4 \pi N e^2 K^2\gamma}{c}\Theta(\alpha_n^2) \times \nonumber \\
	&\sum\limits_{n=1}^{\infty}
	\left[{J_n^\prime}^2(x_n)+\left(\frac{\alpha_n}{K}-\frac{n}{x_n}\right)^2 J_n^2(x_n)\right]\;,
\end{align}
where
\begin{align*}
	K &=\frac{e B \lambda_0}{2\pi n c^2} &
	\alpha_n^2 &= n/r -1-K^2 \\
	x_n &= 2 K r \alpha_n &
	r &=\omega/(2\gamma^2\omega_0)\\
	\omega_0 &= 2\pi \beta^* c/\lambda_0 &
	\beta^* &=\beta \left[1-(K/\gamma)^2\right]^{1/2}
\end{align*}
$n$ is the harmonic number, $\lambda_0$ is the undulator's period, $K$ is the undulator parameter, $N$ is the number of the undulator's periods, $J_n(\cdot ),J_n^\prime(\cdot )$ are the $n$th order Bessel function and its derivative, $\Theta(\cdot )$ is the Heaviside step function.

For the $n$-th harmonics, the maximum frequency coincides with \eqref{eq:maxom} :
\[
\max \omega \equiv \omega_n^* = \frac{2\gamma^2\omega_0 n}{1+K^2}\; .
\]

We introduce a new variable $\zeta$ equal to the ratio of the frequency $\omega$ to the maximum frequency,
\[
\zeta_n=\zeta_n(K)=\frac{\omega (1+K^2)}{2\gamma^2\omega_0 n}\;,
\]
with the support  in interval $[0,1]$.

With this variable, the above definitions read
\begin{align*}
	\omega &= 2 \zeta_n\gamma^2\omega_0 n /(1+K^2)\; ;\\
	\D \omega &= \frac{2\gamma ^2\omega_0 n}{(1+K^2)}\D \zeta_n\; ;\\
	\alpha_n^2 &= (1+K^2)(1-\zeta_n )/\zeta_n \\
	x_n &= 2 K n\left[ \frac{\zeta_n (1-\zeta_n)}{(1+K^2)}\right]^{1/2}\; ; \\
	\Theta (\alpha_n ) &=\Theta (1-\zeta_n)\; ;\\
	\left( \frac{\alpha_n}{K}-\frac{n}{x_n}\right)^2 &= \frac{(1+K^2)(1-2\zeta_n)^2 }{4 K^2 \zeta_n (1-\zeta_n)}\; ; \\
	4{J_n^\prime}^2 &=J_{n-1}^2-2J_{n-1}J_{n+1}+J_{n+1}^2\; .
\end{align*}

Focused on ``quantization'' of the multi-harmonics spectrum, we can retain the terms under the sum sign in \eqref{eq:Kin25}:
\begin{align}\label{eq:nthterm}
	S_n(\zeta_n,K)\equiv &\left[J_{n-1}^2(x_n)-2J_{n-1}(x_n)J_{n+1}(x_n)+J_{n+1}^2(x_n) + \vphantom{ \frac{(1+K^2)(1-2\zeta_n)^2 }{K^2 \zeta_n (1-\zeta_n)}}\right. \nonumber \\
	&\left.
	\frac{(1+K^2)(1-2\zeta_n)^2 }{K^2 \zeta_n (1-\zeta_n)}J_{n}^2(x_n)\right] \Theta (1-\zeta_n) \; .
\end{align}

\begin{figure*} [htb]
	\includegraphics[width=\columnwidth]{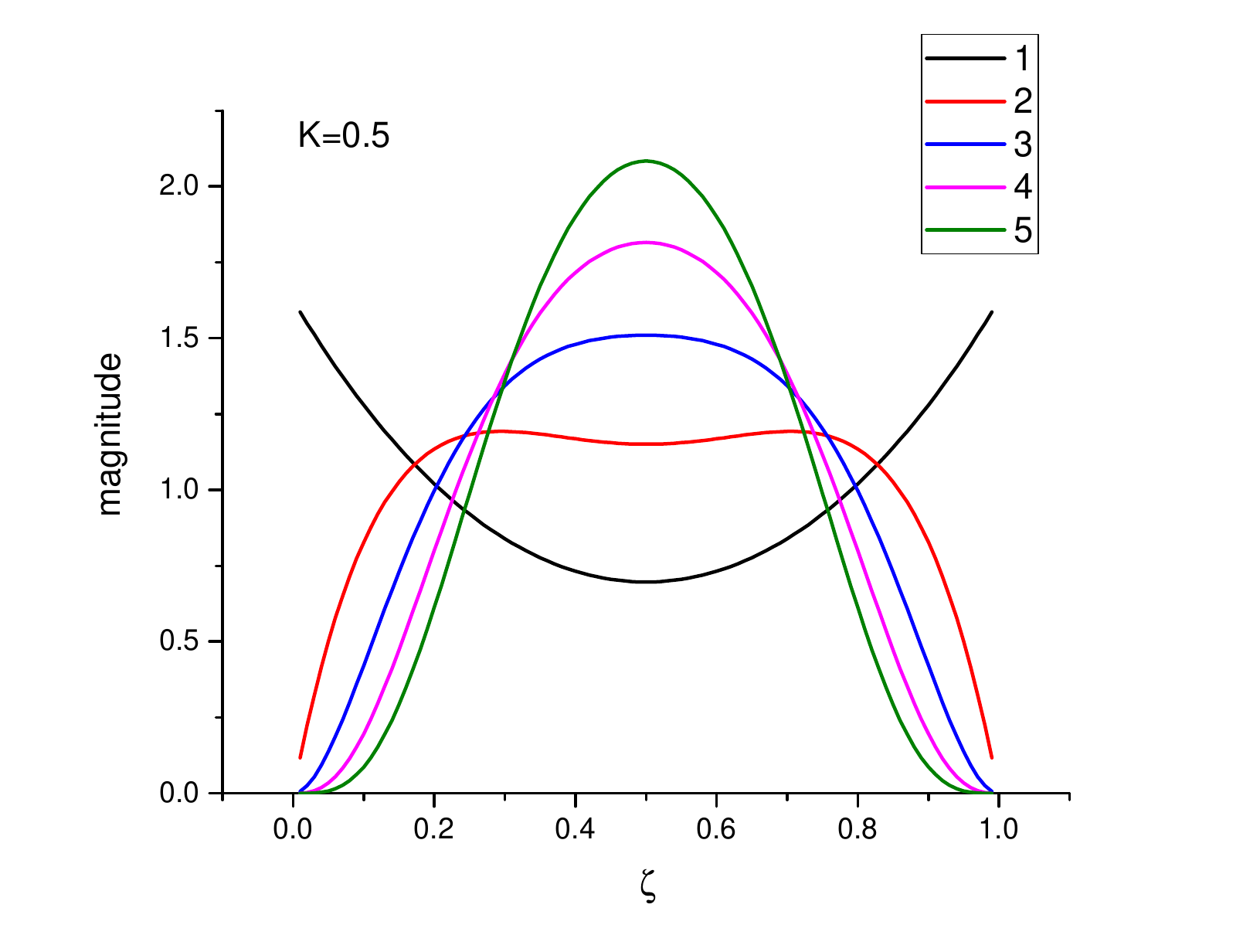}
	\includegraphics[width=\columnwidth]{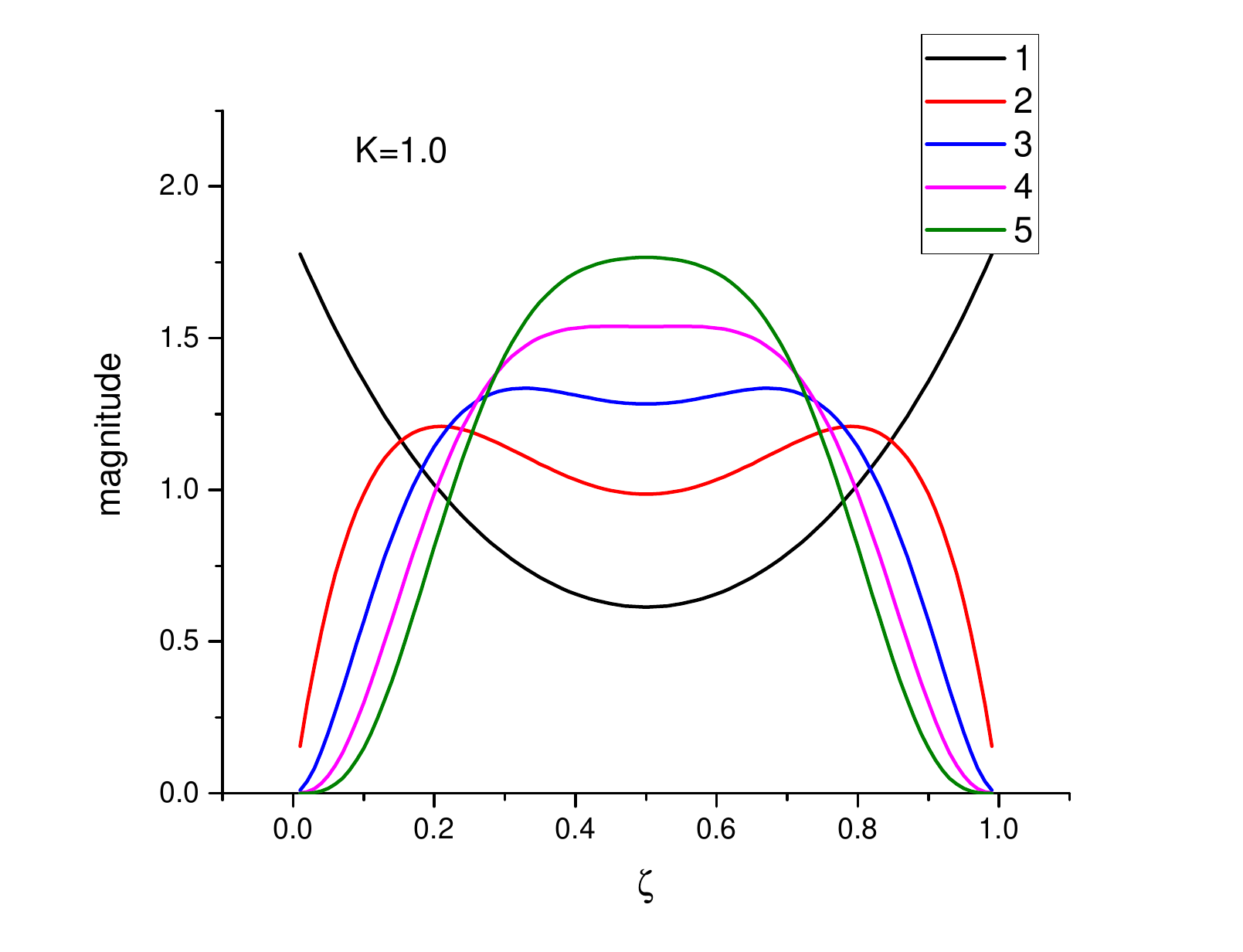}\\
	\includegraphics[width=\columnwidth]{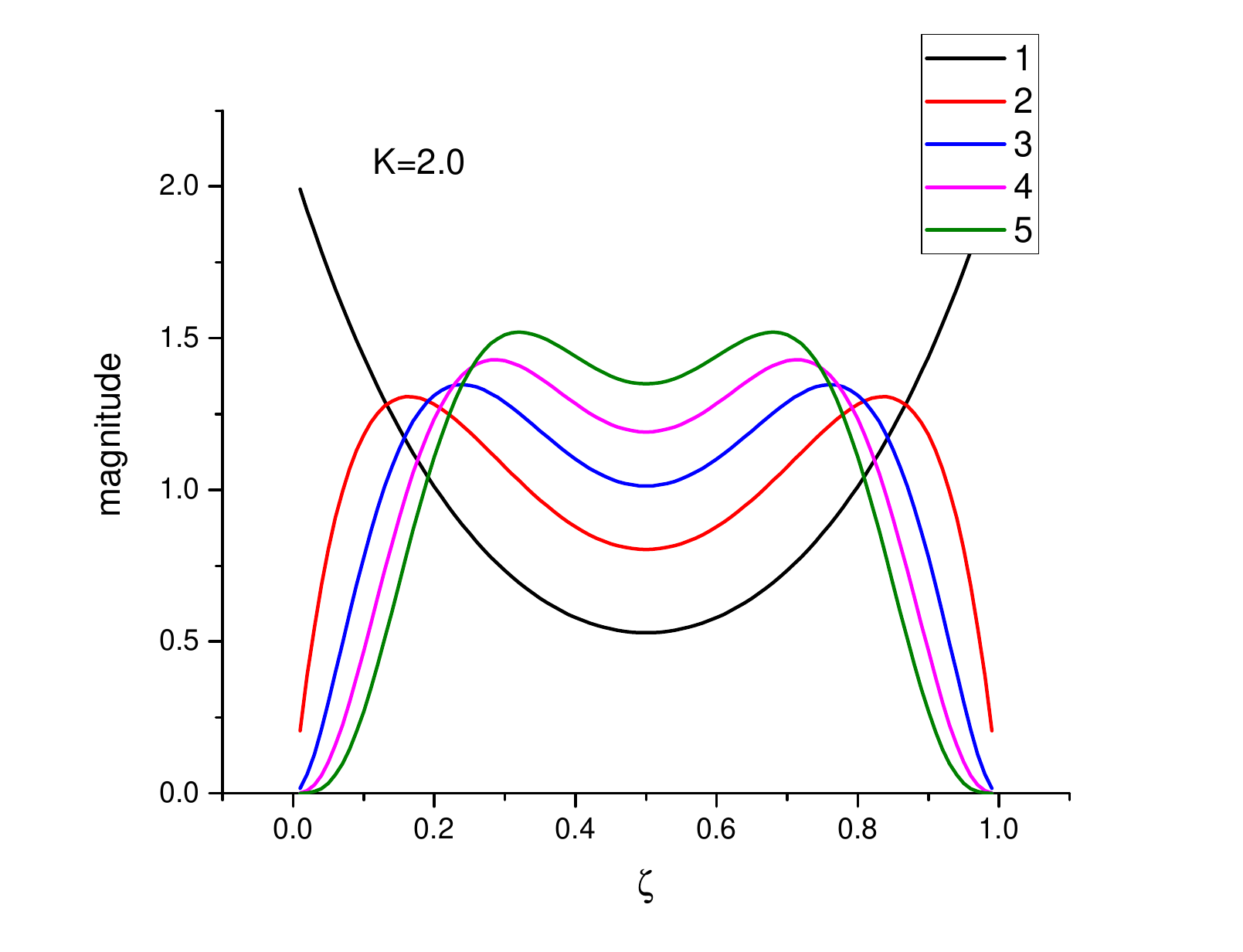}
	\includegraphics[width=\columnwidth]{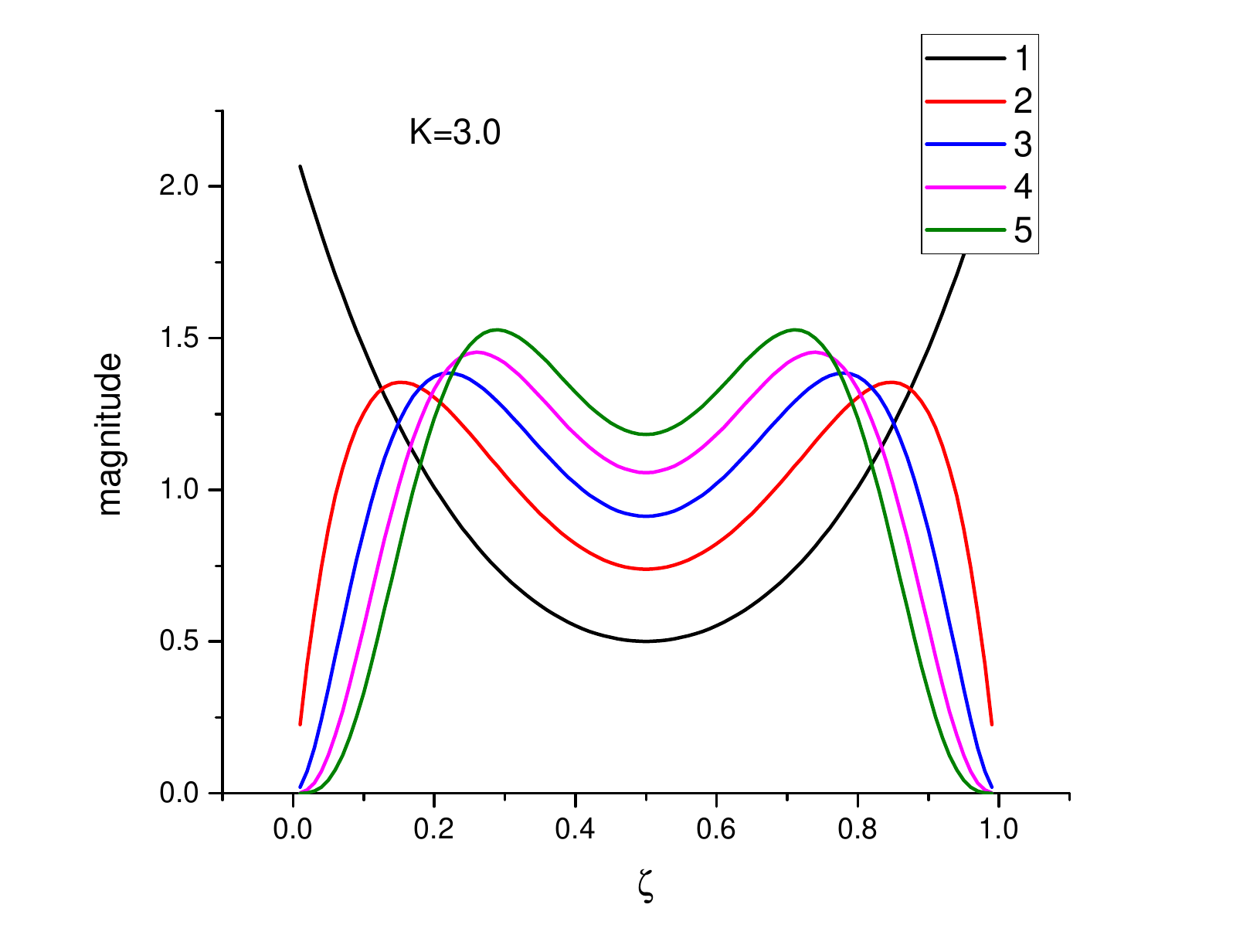}
	\caption{The normalized spectra of the five first harmonics for parameter $K=0.5,1,2,3$. \label{fig:5spectra}}
\end{figure*}

Figure~\ref{fig:5spectra} presents the reduced spectra for the first five harmonics at a different magnitude of $K$.  As presented in Fig.~\ref{fig:5spectra}, the shape of the first harmonic remains about the same for different $K$s, while the higher harmonics shapes vary. All the harmonics are symmetrical around $\zeta = 1/2 $, which confirms its mean value to be equal to a half of the maximum.

\section{Sample of undulator tuning\label{apb}}
We illustrate fine tuning of a hypothetical undulator to increase its efficiency. The sample undulator is equivalent to circularly polarized infrared laser interacting head-on with 1\,GeV electrons, analogous to the alternative source of polarized positrons for ILC  \cite{ilcrdr}. The source employs the first harmonic at $\lambda \to 0$.

\begin{figure*} 
	\includegraphics[width=\textwidth]{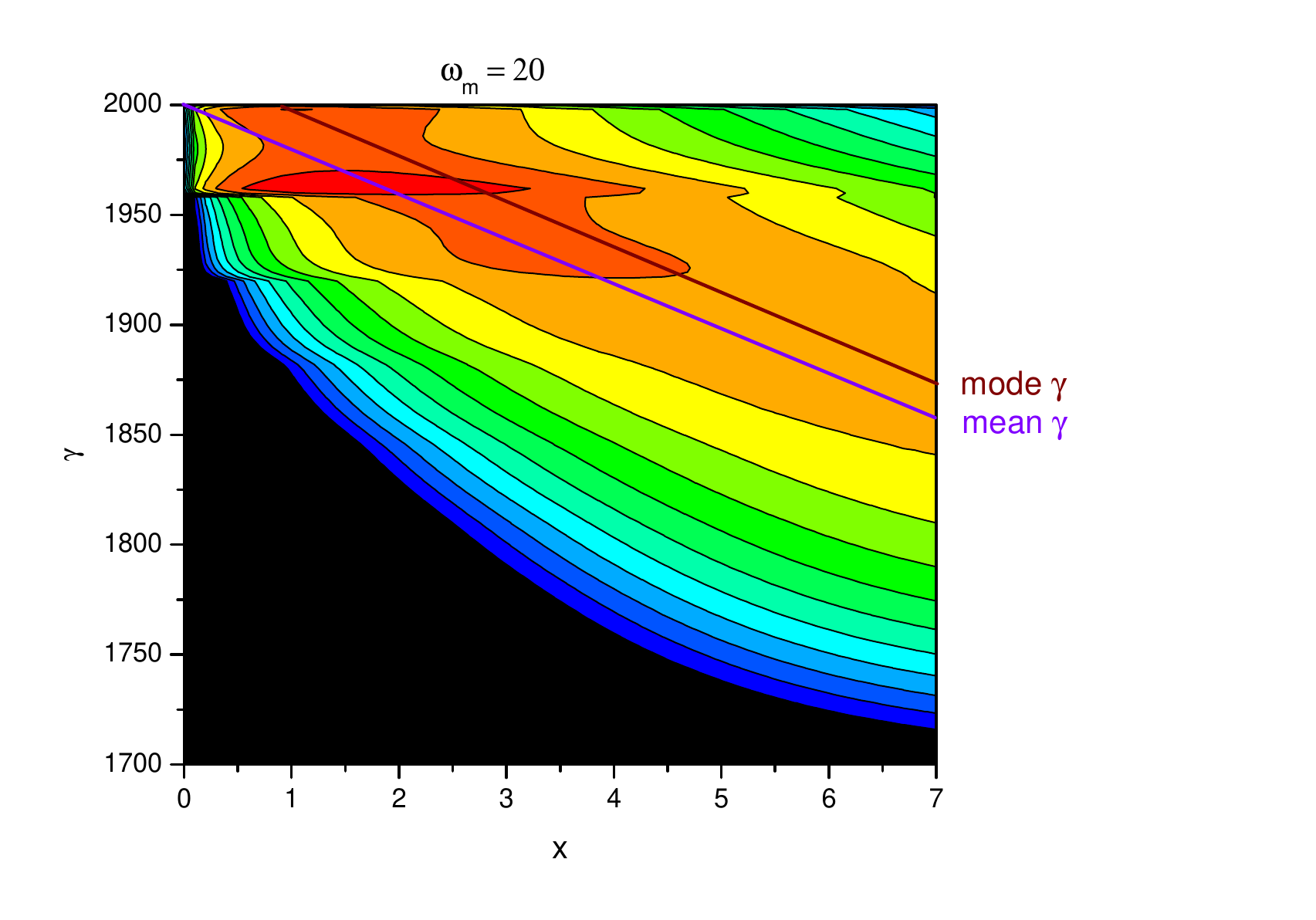}
	\caption{Evolution of the spectrum in a hypothetical case of 1~GeV electrons scattered off 1~eV photons, $\lambda \to 0$ (dipole harmonic). \label{fig:meanmode}}
\end{figure*}

Evolution of the initially delta-shaped electron spectrum is presented in Fig.~\ref{fig:meanmode}.  An optimal tuning may be the following:  for the initial interval $0 < x \lesssim 1$ the resonant frequency should adjust to the maximum in the initial electron spectrum.  Then, for the interval $1\lesssim  x \lesssim 3$ the resonant frequency is adjusted to $E_\text{init} - \omega_\text{max}$.  The remaining fraction of the undulator length should follow the mode line with a constant shift to the right from the mean line.  

Such adjustment will produce  10\dots30\,\% brighter radiation compared with adjustment to the mean frequency.   
The proposed tuning qualitatively agrees with an  empirical approach described in \cite{mak16}.

\end{document}